\documentclass[
reprint,
amsmath,amssymb,
aps,
]{revtex4-2}

\usepackage{graphicx}
\usepackage{dcolumn}
\usepackage{bm}

\usepackage{amsthm}
\newtheorem{remark}{Remark}

\usepackage{braket}
\usepackage{caption}
\usepackage{xcolor}
\usepackage{diagbox}
\usepackage{booktabs}
\usepackage{adjustbox}
\usepackage{multirow}
\usepackage[mathlines]{lineno}
\usepackage{ragged2e}
\usepackage{pifont}
\usepackage{makecell}
\usepackage{algorithm}
\usepackage{algpseudocode}
\usepackage{soul}
\soulregister\ensuremath{1}
\renewcommand{\hl}[1]{#1} 

\newcommand{\boldb}{\mathbf{b}}

\newcommand{\boldp}{\mathbf{p}}
\newcommand{\boldv}{\mathbf{v}}

\newcommand{\E}{\mathbb{E}}
\newcommand{\R}{\mathbb{R}}
\newcommand{\tr}{\text{tr}}

\newcommand{\tcb}[1]{\textcolor{blue}{#1}}

\newcommand{\cmark}{\textcolor{green!70!black}{\ding{51}}}
\newcommand{\xmark}{\textcolor{red}{\ding{55}}}

\usepackage[most]{tcolorbox}
\newtcolorbox{myhighlight}{
    colback=yellow!30, 
    frame hidden,
    boxrule=0pt,
    sharp corners,
    enhanced,
    breakable 
}

\begin{document}


\title{\textbf{Operator-aware shadow importance sampling for accurate fidelity estimation}}

\author{Hyunho Cha${}^{1}$}
\email{ovalavo@snu.ac.kr}
\author{Sangwoo Hong${}^{2,3}$}
\email{swhong06@konkuk.ac.kr}
\author{Jungwoo Lee${}^{1,3}$}
\email{junglee@snu.ac.kr} 
\affiliation{${}^{1}$NextQuantum and Department of Electrical and Computer Engineering, Seoul National University, Seoul 08826, Republic of Korea}
\affiliation{${}^{2}$Department of Computer Science and Engineering, Konkuk University, Seoul 05029, Republic of Korea}   
\affiliation{${}^{3}$(Corresponding authors: Sangwoo Hong; Jungwoo Lee)}

\date{\today}

\begin{abstract}
Estimating the fidelity between an unknown quantum state and a fixed target is a fundamental task in quantum information science. Direct fidelity estimation (DFE) enables this without full tomography by sampling observables according to a target-dependent distribution. However, existing approaches face notable trade-offs. Grouping-based DFE achieves strong accuracy for small systems but suffers from exponential scaling, and its applicability is restricted to Pauli measurements. In contrast, classical-shadow-based DFE offers scalability but yields lower accuracy on structured states. In this work, we address these limitations by developing two classes of \emph{operator-aware shadow importance sampling} algorithms using informationally overcomplete positive operator-valued measures. Instantiated with local Pauli measurements, our algorithm improves over the grouping-based algorithms for Haar-random states. For structured states such as the GHZ and W states, our algorithm also eliminates the exponential memory requirements of previous grouping-based methods. Numerical experiments confirm that our methods achieve state-of-the-art performance across Haar-random, GHZ, and W targets.
\end{abstract}

\maketitle


\section{Introduction}

\begin{figure}
    \centering
    \includegraphics[width=\linewidth]{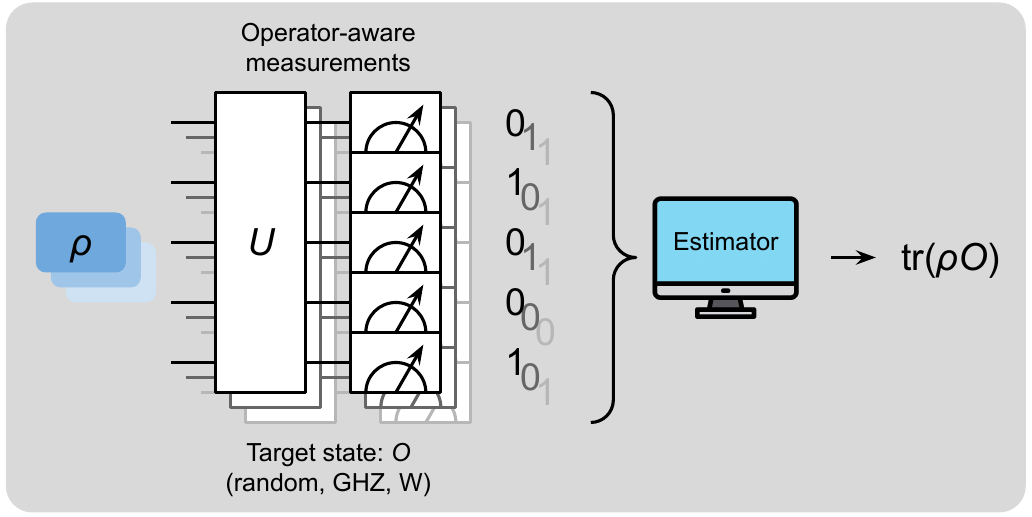}
    \caption{\justifying General framework of DFE. The goal is to estimate \(\tr(\rho O)\) for a target state \(O\). Given many copies of the unknown state \(\rho\), random measurements are performed according to a distribution optimized for each \(O\). After collecting the measurement statistics, the estimation algorithm produces an estimate of the fidelity.}
    \label{figure:scheme}
\end{figure}

As quantum processors grow in size and complexity, efficient verification tools become crucial for assessing device performance \cite{knill2008randomized, barz2013experimental, fitzsimons2018post, erhard2019characterizing, gheorghiu2019verification, wright2019benchmarking, blume2020volumetric, helsen2022general, polloreno2025theory}. Among such tasks, estimating the fidelity between an unknown state and a fixed target is a core task in quantum information \cite{flammia2011direct, cerezo2020variational, zhang2021direct, wang2022quantum, qin2024experimental, seshadri2024theory}. In contrast to target-agnostic approaches that first collect measurement data independent of the target and subsequently perform operator-specific postprocessing \cite{d2007optimal, huang2020predicting, huang2022learning, caprotti2024optimizing, mangini2025low}, target-aware methods optimize the measurement distribution to minimize estimation error \cite{garcia2021learning, huang2021efficient, yen2023deterministic}. Direct fidelity estimation (DFE) provides an efficient approach to this task by sampling observables from a distribution tailored to the target, avoiding full tomography and often yielding dramatically fewer measurements (as indicated in Fig.~\ref{figure:scheme}). Among recent developments in this field, two approaches have demonstrated significant efficacy. First, DFE with grouping Pauli operators~\cite{barbera2025sampling} (referred to as G-DFE in this work) exploits qubit-wise commutativity (QWC) to estimate many Pauli expectations from a single local measurement setting. It directly extends \cite{flammia2011direct} and achieves the best accuracy on Haar-random states, GHZ states \cite{greenberger1989going}, and W states \cite{dur2000three, cabello2002bell}. However, its grouping procedure scales exponentially with the system size, making it impractical for larger systems. Moreover, G-DFE is inherently limited to Pauli measurements and cannot be directly applied to more general measurement settings. Second, the classical-shadow-based DFE~\cite{cha2025efficient} (referred to as C-DFE in this work) leverages the structure of specific targets to optimize the sampling distribution while retaining efficient sampling and postprocessing. It is a scalable algorithm and consistently outperforms the original DFE in~\cite{flammia2011direct}. For small-scale systems, however, its estimation accuracy is slightly lower than that of G-DFE.

In this paper, we address the aforementioned limitations. Our contribution is two-folded:

\begin{itemize}
\item First, we develop an operator-aware importance sampler that operates with \emph{any} informationally overcomplete positive operator-valued measure (IOC-POVM) obtained by solving a linear program (LP) over IOC-POVM expansions of the target. Specifically, we expand the target operator in an overcomplete \(6^n\)-element POVM, introducing free parameters that can be tuned to minimize estimator variance. In contrast, G-DFE operates within the \(4^n\) Pauli basis, whose coefficients are uniquely determined and thus offer no comparable optimization freedom. Instantiated with local Pauli measurements, our algorithm surpasses G-DFE, the previous state-of-the-art approach, for Haar-random targets.

\item Moreover, we develop an operator-aware importance sampler for highly structured targets, where grouping is more effective, such as the GHZ and W states.
We introduce a scalable grouping-based approach that handles such targets.
In this way, the proposed estimators inherit C-DFE's efficiency while matching (GHZ) or surpassing (W) the accuracy of G-DFE.
\end{itemize}

The remainder of this paper is organized as follows. Section~\ref{section:notations_and_prelim} introduces the notations and provides prior DFE protocols as preliminaries. Our proposed optimization framework for arbitrary states is presented in Section~\ref{section:General}, followed by our proposed optimization framework for the GHZ and W states in Section~\ref{section:special}. Numerical results are reported in Section~\ref{section:experiments}. Finally, Section~\ref{section:discussion} concludes the paper with discussion and outlook.

\section{Notations and preliminaries}
\label{section:notations_and_prelim}
\subsection{Notations}
Let \(n\) denote the number of qubits, and set \(d = 2^n\) for the dimension of the associated Hilbert space.  
For a binary vector \(\boldb\), we write \(|\boldb|\) for its Hamming weight.  
For a single-qubit Pauli operator \(P \in \{I, X, Y, Z\}\) and an \(n\)-qubit Pauli string \(\boldp \in \{I, X, Y, Z\}^n\), we define \(\boldp_P\in\{0,1\}^n\) such that
\[
(\boldp_P)_i = \begin{cases}
1 & \boldp_i = P\\
0 & \text{otherwise}
\end{cases}.
\]
For a vector \(\mathbf{x}\in\R\) and an index set \(\mathcal{I}\), we denote by \(\mathbf{x}_\mathcal{I} \in \R^{|\mathcal{I}|}\) the subvector of \(\mathbf{x}\) consisting of the entries indexed by \(\mathcal{I}\). \(\mathbf{1}_A(x)=[x\in A]\) denotes the indicator function. Bold symbols \(\mathbf{0}\) and \(\mathbf{1}\) denote the all-zeros and all-ones vectors, respectively. For integers \(a\) and \(b\) with \(b\ge a\), \(\mathrm{unif}\{a,b\}\) denotes the discrete uniform distribution over \(\{a,a+1,\dots,b-1,b\}\).

\subsection{Preliminaries on previous DFE protocols}
We briefly describe the DFE algorithm based on sampling Pauli operators introduced in~\cite{flammia2011direct}. For any Hermitian operator \(A\), its \emph{characteristic function} is defined as
\[
\chi_A(\boldp) = \tr\!\left(\frac{A \bigotimes_{i=1}^n \boldp_i}{\sqrt{d}}\right),
\]
that is, the normalized expectation value of the Pauli string \(\boldp\). We denote the corresponding characteristic vector, indexed by Pauli strings, as \(\boldsymbol\chi_A\). Suppose the unknown state is \(\rho\), and let the target pure state have density matrix \(O\). Then the fidelity can be expressed as
\[
\tr(\rho O) = \sum_\boldp \chi_\rho(\boldp) \chi_O(\boldp).
\]
Then, this quantity can be estimated as follows. First, select \(\boldp\) at random with probability
\begin{equation}
\label{equation:dfe_probabilities}
\tilde{p}(\boldp) = \chi_O(\boldp)^2.
\end{equation}
Since \(O\) is pure, this indeed yields a normalized probability distribution. Then, define the random variable
\begin{equation}
\label{equation:flammia_random_variable_definition}
\tilde{R} = \frac{\chi_\rho(\boldp)}{\chi_O(\boldp)}.
\end{equation}
It is straightforward to verify that \(\E[\tilde{R}]=\tr(\rho O)\). However, since the numerator in \eqref{equation:flammia_random_variable_definition} is unknown, the random variable \(\tilde{R}\) must be estimated from repeated measurements.

G-DFE builds on this method by grouping qubit-wise commuting Pauli operators, thereby allowing the simultaneous estimation of multiple Pauli expectation values from a single measurement setting. Concretely, a measurement setting \(\boldp\) is sampled from a subset of \(\{I,X,Y,Z\}^n\) according to the distribution \(p(\boldp)\), where \(\boldp\) represents a group of Pauli operators that (qubit-wise) commute with \(\boldp\). Then \(\mathcal{I}(\boldp)\) is defined as the index set of the characteristic vector corresponding to the Pauli strings in the sampled group. In G-DFE, the probability of sampling \(\boldp\) is given by \(\left\|(\boldsymbol\chi_O)_{\mathcal{I}(\boldp)}\right\|^2\). Then the random variable associated with \(\boldp\) is defined as
\begin{equation}
\label{equation:barbera_random_variable_definition}
R = \frac{(\boldsymbol\chi_\rho)_{\mathcal{I}(\boldp)} \cdot (\boldsymbol\chi_O)_{\mathcal{I}(\boldp)}}{\left\|(\boldsymbol\chi_O)_{\mathcal{I}(\boldp)}\right\|^2} = \frac{(\boldsymbol\chi_\rho)_{\mathcal{I}(\boldp)} \cdot (\boldsymbol\chi_O)_{\mathcal{I}(\boldp)}}{\tilde{p}(\boldp)}.
\end{equation}
In order to reduce variance, G-DFE employs the \emph{sorted insertion} (SI) algorithm \cite{crawford2021efficient} to group Pauli operators, as outlined in Algorithm~\ref{algorithm:sorted_insertion_algorithm} in Appendix~\ref{appendix:sorted_insertion}. However, by construction, this approach is limited to local Pauli measurements (whether grouping is used or not) and cannot leverage more general POVMs.

\section{Optimization for general states}
\label{section:General}
In this section, we propose an optimization algorithm that can be applied to any IOC-POVM, namely \emph{OASIS-GT} (operator-aware shadow importance sampling for general targets). The key idea is to expand the target operator in an IOC-POVM, which introduces non-unique coefficients. We keep the estimator unbiased, and choose the sampling law so that the worst-case variance is minimized. The proposed estimation procedure for general states is outlined in Algorithm~\ref{algorithm:oasis_lp}, and its complete derivation is provided in Appendix~\ref{appendix:algorithm_1_derivation}.

\setcounter{figure}{0}
\renewcommand{\figurename}{ALGORITHM}
\begin{figure}[t]
\centering
\begin{minipage}{\linewidth}
\caption{\justifying OASIS-GT.}
\label{algorithm:oasis_lp}
\rule{\linewidth}{1pt}
\begin{algorithmic}[0]
\State \underline{\textbf{Estimator optimization}}
\State \textbf{Input:} POVM $\boldsymbol{\Pi} = \{\Pi_{U,\boldb}\}_{U,\boldb}$ and default distribution $p$
\State \textbf{Output:} Weights $\omega$
\State Solve the following LP:\\
$\displaystyle
\begin{aligned}
\mathop{\text{minimize}}_{\omega,t} \quad & \sum_U p(U) t_U\\
\text{subject to} \quad & -t_U \le \omega_{U,\boldb} \le t_U, \quad \forall U,\boldb,\\
& \sum_{U,\boldb} \omega_{U,\boldb} \Pi_{U,\boldb} = O.
\end{aligned}
$
\State \textbf{return} \(\{\omega_{U,\boldb}\}_{U,\boldb}\)\\
\State \underline{\textbf{Estimation}}
\State \textbf{Input:} State $\rho$, weights $\omega$, default distribution $p$,
\State and number of shots $N$
\State \textbf{Output:} Estimate of $\tr(\rho O)$, where
\State $O = \sum_{U,\boldb} \omega_{U,\boldb} p(U)U^\dagger|\boldb\rangle\langle\boldb|U$
\State $\text{sum} \gets 0$
\For{\underline{\hspace{0.5em}} in range($N$)}
  \State Sample \(U \sim q(U) = \frac{p(U) \max_\boldb |\omega_{U,\boldb}|}{\sum_{U^\prime} p(U^\prime) \max_\boldb |\omega_{U^\prime,\boldb}|}\).
  \State $\rho^\prime \gets U \rho U^\dagger$
  \State Measure $\rho^\prime$ in the computational basis and get $\boldb$.
  \State \(S(U,\boldb) \gets \omega_{U,\boldb} p(U) / q(U)\);\quad\(\text{sum} \, \texttt{+=} \, S(U,\boldb)\)
\EndFor
\State \textbf{return} $\text{sum} / N$
\end{algorithmic}
\rule{\linewidth}{1pt}
\end{minipage}
\end{figure}

OASIS-GT provides a general solution for random targets. However, as will be demonstrated in Section~\ref{section:experiments}, applying OASIS-GT to highly structured states such as the GHZ and W states shows diminished performance when compared with previous algorithms such as G-DFE and C-DFE. This is because for these special states, the Pauli operators with nonzero probabilities are sparse and can be grouped efficiently, which eliminates the potential advantage of optimizations that struggle to recognize this property (for a more detailed discussion, see Remark~\ref{remark:why_lp_worse} in Appendix~\ref{appendix:algorithm_1_derivation}).
\hl{In this regard, we remark that OASIS-GT is \emph{not} universally preferable. While it achieves the best performance for Haar-random states, which exhibit minimal structure, we leave a complete characterization of the regimes in which it is most effective to future work.}

\begin{figure}[t]
\centering
\begin{minipage}{\linewidth}
\caption{\justifying \hl{OASIS-GHZ}.}
\label{algorithm:oasis_st_ghz}
\rule{\linewidth}{1pt}
\begin{algorithmic}[0]
\State \textbf{Input:} State $\rho$ and \((\epsilon,\delta)\)
\State \textbf{Output:} Estimate of $\tr(\rho O)$, where \(O\) is the \(n\)-qubit
\State GHZ state density matrix
\State $(U_{X}, U_Y, U_Z) \gets (H, HS^\dagger, I)$
\State \(l \gets \left\lceil \frac{1}{\epsilon^2 \delta} \right\rceil\);\quad\(m \gets \left\lceil \frac{2}{l\epsilon^2}\ln\frac{2}{\delta} \right\rceil\);\quad$\text{sum} \gets 0$
\For{\underline{\hspace{0.5em}} in range($l$)}
    \State \(\hat{R} \gets 0\);\quad Flip a fair coin.
    \If{heads}
        \State Sample \(k \sim \mathrm{unif}\{1,d\}\).
        \If{\(k\le2\)}\Comment{Branch 0}
            \For{\underline{\hspace{0.5em}} in range($m$)}
                \State \(S \gets 1\);\quad\(\hat{R} \, \texttt{+=} \, S\)
            \EndFor
        \Else\Comment{Branch 1}
            \For{\underline{\hspace{0.5em}} in range($m$)}
                \State Measure \(\rho\) in the Pauli \(Z\) basis and get \(\boldb\).
                \State \(S \gets \frac{d(\delta_{\boldb,\mathbf{0}}+\delta_{\boldb,\mathbf{1}}) - 2}{d-2}\);\quad\(\hat{R} \, \texttt{+=} \, S\)
            \EndFor
        \EndIf
    \Else\Comment{Branch 2}
        \For{\underline{\hspace{0.5em}} in range($m$)}
            \State Sample \(\boldp \in \{X,Y\}^n\) with \(|\boldp_Y| \equiv 0 \, (\mathrm{mod} \, 2)\)
            \State uniformly at random.
            \State \(U \gets \bigotimes_{i=1}^n U_{\boldp_i}\);\quad\(\rho^\prime \gets U\rho U^\dagger\)
            \State Measure \(\rho^\prime\) in the Pauli \(Z\) basis and get \(\boldb\).
            \State \(S \gets (-1)^{|\boldp_Y|/2 + |\boldb|}\);\quad\(\hat{R} \, \texttt{+=} \, S\)
        \EndFor
    \EndIf
    \State \(\hat{R} \, \texttt{/=} \, m\);\quad\(\text{sum} \, \texttt{+=} \, \hat{R}\)
\EndFor
\State \textbf{return} \(\text{sum}/l\)
\end{algorithmic}
\rule{\linewidth}{1pt}
\end{minipage}
\end{figure}

\begin{figure}[t]
\centering
\begin{minipage}{\linewidth}
\caption{\justifying \hl{OASIS-W}.}
\label{algorithm:oasis_st_w}
\rule{\linewidth}{1pt}
\begin{algorithmic}[0]
\State \textbf{Input:} State $\rho$ and \((\epsilon,\delta)\)
\State \textbf{Output:} Estimate of $\tr(\rho O)$, where \(O\) is the \(n\)-qubit
\State W state density matrix
\State $(U_{X}, U_Y, U_Z) \gets (H, HS^\dagger, I)$
\State \(l \gets \left\lceil \frac{1}{\epsilon^2 \delta} \right\rceil\);\quad\(m_1 \gets \left\lceil \frac{2 n^2}{l\epsilon^2} \left(\frac{2\binom{n-1}{\lfloor n/2\rfloor} - 1}{d-n}\right)^2 \ln\frac{2}{\delta} \right\rceil\);
\State \(m_2 \gets \left\lceil \frac{n^2}{2l\epsilon^2} \ln\frac{2}{\delta} \right\rceil\);\quad$\text{sum} \gets 0$\\
\For{\underline{\hspace{0.5em}} in range($l$)}
    \State \(\hat{R} \gets 0\);\quad Sample \(k \sim \mathrm{unif}\{1,n\}\).
    \If{\(k=1\)}
        \State Sample \(k \sim \mathrm{unif}\{1,d\}\).
        \If{\(k\le n\)}\Comment{Branch 0}
            \State \(m \gets 1\);\quad\(\hat{R} \, \texttt{+=} \, 1\)
        \Else\Comment{Branch 1}
            \State \(m \gets m_1\)
            \For{\underline{\hspace{0.5em}} in range($m$)}
                \State Measure \(\rho\) in the Pauli \(Z\) basis and get \(\boldb\).
                \State \(S \gets \frac{d \mathbf{1}_{\{1\}}(|\boldb|) - n}{d-n}\);\quad\(\hat{R} \, \texttt{+=} \, S\)
            \EndFor
        \EndIf
    \Else\Comment{Branch 2}
        \State \(m \gets m_2\)
        \For{\underline{\hspace{0.5em}} in range($m$)}
            \State Sample \(\boldp=\boldp^{(X/Y,i,j)}\) uniformly at random.
            \State \(U \gets \bigotimes_{i=1}^n U_{\boldp_i}\);\quad\(\rho^\prime \gets U\rho U^\dagger\)
            \State Measure \(\rho^\prime\) in the Pauli \(Z\) basis and get \(\boldb\).
            \State \(S \gets \frac{n}{2} (-1)^{b_i + b_j} \mathbf{1}_{\{0\}}\left(\left|\boldb_{[n]\setminus\{i,j\}}\right|\right)\);\quad\(\hat{R} \, \texttt{+=} \, S\)
        \EndFor
    \EndIf
    \State \(\hat{R} \, \texttt{/=} \, m\);\quad\(\text{sum} \, \texttt{+=} \, \hat{R}\)
\EndFor
\State \textbf{return} \(\text{sum}/l\)
\end{algorithmic}
\rule{\linewidth}{1pt}
\end{minipage}
\end{figure}

\section{Optimization for structured states}
\label{section:special}

In this section, we present efficient and scalable algorithms for structured states, specifically the GHZ and W states. Heuristic algorithms such as G-DFE show scalability issues due to the exponential resource requirements in the grouping procedure and the memory needed to store the groups. To overcome these limitations, we propose more efficient estimators by leveraging non-heuristic grouping strategies.
\hl{While this work focuses on algorithmic instantiations for GHZ and W states, the underlying optimization framework is general in spirit. Specifically, it relies on the existence of a compact description of the Pauli operators that carry nonzero weight, together with an explicit partition into QWC groups. We expect that analogous algorithms can be developed for other classes of structured states, albeit with additional problem-specific design efforts.}

\subsection{\hl{OASIS-GHZ}}
\label{section:oasis_st_ghz}

The proposed estimation procedure for the GHZ state is outlined in Algorithm~\ref{algorithm:oasis_st_ghz}, and its complete derivation is provided in Appendix~\ref{appendix:algorithm_2_derivation}. The variance of this estimator can be tightly bounded in terms of the true fidelity, and in particular, it vanishes as the fidelity approaches 1 (see Appendix~\ref{appendix:algorithm_2_variance}).

\subsection{\hl{OASIS-W}}

The proposed estimation procedure for the W state is outlined in Algorithm~\ref{algorithm:oasis_st_w}, and its complete derivation is provided in Appendix~\ref{appendix:algorithm_3_derivation}.

\subsection{\hl{Improvement for the W state}}

Fewer groups generally lead to smaller mean squared error (MSE), since more Pauli observables can be estimated from each measurement shot. In this sense, Algorithms~\ref{algorithm:oasis_st_ghz} and \ref{algorithm:oasis_st_w} have the desirable property of minimizing the number of groups.

For the W state, the grouping obtained by G-DFE is not only inefficient but also suboptimal. For example, in the 4-qubit W state, the following six strings are grouped together in G-DFE:
\[
YYII, \, YYIZ, \, YIYI, \, YIYZ, \, IYYI, \, IYYZ.
\]
This grouping is valid, since all of these strings commute with \(YYYZ\), but it is suboptimal and redundant because \(YYYZ\) itself has zero probability. Consequently, this grouping does not minimize the number of groups, as shown in Table~\ref{table:num_groups} in Appendix~\ref{section:additional_tables}. On the other hand, \hl{OASIS-W (and OASIS-GHZ)} minimizes the number of groups while ensuring that each group contains no redundant Pauli strings, which accounts for the improvement achieved by our method.

\section{Numerical results}
\label{section:experiments}

\subsection{Setting}

We conducted numerical experiments on Haar-random, GHZ, and W states for systems of 3, 4, 5, and 6 qubits to demonstrate the performance of the proposed algorithms. For all targets, we evaluated the proposed OASIS-GT method. \hl{In addition, we applied OASIS-GHZ to the GHZ states and OASIS-W to the W states.} In each experiment, the target density matrix is denoted by \(O\), and the unknown state is modeled as a depolarized version
\[
\rho = (1-p)O+p\frac{I}{d}
\]
with \(p=0.1\).

\subsection{{MSE comparison}}
\label{section:mse_comparison}

\begin{table}[t!]
\caption{\label{table:haar_experiments}\justifying Comparison of MSE (1e-4) for Haar-random states. C-DFE and \hl{OASIS-GHZ/W} are inapplicable, as they are designed for structured targets.}
\centering
\begin{tabular}{l|cccc}
\toprule
\multicolumn{1}{c|}{$n$} & 3 & 4 & 5 & 6\\
\midrule
G-DFE & 4.34 & 3.07 & 2.91 & 2.23\\ \hline
\textbf{OASIS-GT} & \textbf{\tcb{3.56}} & \textbf{\tcb{2.39}} & \textbf{\tcb{2.00}} & \textbf{\tcb{1.53}}\\
\bottomrule
\end{tabular}
\end{table}

\begin{table}[t!]
\caption{\label{table:ghz_experiments}\justifying Comparison of MSE (1e-4) for the GHZ state.}
\centering
\begin{tabular}{l|cccc|c}
\toprule
\multicolumn{1}{c|}{$n$} & 3 & 4 & 5 & 6 & Scalability\\
\midrule
G-DFE & \textbf{\tcb{1.30}} & \textbf{\tcb{1.01}} & \textbf{\tcb{.953}} & \textbf{\tcb{.954}} & \xmark\\ \hline
\textbf{OASIS-GT} & 1.77 & 1.72 & 1.56 & 1.51 & \xmark\\ \hline
C-DFE & 1.45 & 1.46 & 1.44 & 1.45 & \cmark\\ \hline
\textbf{\hl{OASIS-GHZ}} & \textbf{\tcb{1.30}} & \textbf{\tcb{1.01}} & \textbf{\tcb{.953}} & \textbf{\tcb{.954}} & \cmark\\
\bottomrule
\end{tabular}
\end{table}

\begin{table}[t!]
\caption{\label{table:w_experiments}\justifying Comparison of MSE (1e-4) for the W state.}
\centering
\begin{tabular}{l|cccc|c}
\toprule
\multicolumn{1}{c|}{$n$} & 3 & 4 & 5 & 6 & Scalability\\
\midrule
G-DFE & 2.77 & 2.58 & 1.68 & .843 & \xmark\\ \hline
\textbf{OASIS-GT} & 2.63 & 3.66 & 4.60 & 5.31 & \xmark\\ \hline
C-DFE & \textbf{\tcb{2.16}} & 3.02 & 3.08 & 2.78 & \cmark\\ \hline
\textbf{\hl{OASIS-W}} & 2.77 & \textbf{\tcb{2.35}} & \textbf{\tcb{1.40}} & \textbf{\tcb{.721}} & \cmark\\
\bottomrule
\end{tabular}
\end{table}

The results are summarized in Tables~\ref{table:haar_experiments}, \ref{table:ghz_experiments}, and \ref{table:w_experiments} for Haar-random, GHZ, and W states, respectively. All reported values are averaged over 1000 trials.

In Tables~\ref{table:ghz_experiments}~and~\ref{table:w_experiments}, the notion of ``scalability'' refers to the asymptotic scaling of the overall algorithmic cost with respect to the number of qubits \(n\). In particular, the symbols \cmark\ and \xmark\ indicate whether the corresponding method admits polynomial-time or exponential-time scaling in \(n\), respectively, taking into account both measurement and classical processing overhead (the polynomial measurement overhead for GHZ/W states follows from the bounds derived in \cite{flammia2011direct}).
For the GHZ case in Table~\ref{table:ghz_experiments}, although G-DFE and OASIS-GHZ achieve the same performance due to identical grouping structures, they differ substantially in classical computational cost. Specifically, G-DFE is marked as non-scalable, whereas OASIS-GHZ is marked as scalable. Further details on this distinction are provided in Remark~\ref{remark:ghz_not_scalable} in Appendix~\ref{appendix:algorithm_2_derivation}.

For OASIS-GT and C-DFE, the user can directly specify the number of measurement shots, and in all reported settings this number is matched to that of G-DFE (see Table~\ref{table:num_shots} in Appendix~\ref{section:additional_tables}) for a fair comparison. In contrast, for \hl{OASIS-GHZ/W}, the user specifies \(\epsilon\) and \(\delta\); the total number of shots then becomes a random variable, and is comparable on average to that of G-DFE. Since the MSE typically scales inversely with the number of shots, we apply a correction factor so that the reported MSEs for \hl{OASIS-GHZ/W} reflect the same average number of shots as G-DFE.

Optimization for Haar-random states was performed using an IOC-POVM based on uniform Pauli measurements, ensuring a consistent basis with G-DFE, which also employs Pauli measurements. The results show that for Haar-random states, OASIS-GT consistently outperforms G-DFE in terms of MSE.
\hl{For the GHZ states, OASIS-GHZ achieves the best performance for systems of three qubits or larger, while for the W states, OASIS-W achieves the best performance for systems of four qubits or larger.}
More importantly, the key advantage of \hl{OASIS-GHZ/W} lies in its scalability. That is, it provides a grouping-based estimator that remains practical for large systems.

\section{Discussion}
\label{section:discussion}

We have introduced a framework for DFE based on operator-aware importance sampling. By formulating estimator optimization as a linear program over an IOC-POVM, OASIS-GT extends and improves upon existing DFE approaches. Furthermore, for structured targets such as the GHZ and W states, the proposed \hl{OASIS-GHZ and OASIS-W} provide improved fidelity estimation without exponential storage overhead.

A limitation of OASIS-GT is that, like G-DFE, it does not scale to many qubits for Haar-random states because most random states lack the structure necessary for an optimal DFE protocol to be formulated and implemented efficiently.
\hl{Although LPs are polynomial-time solvable in problem size, the exponential growth of the problem dimension makes the overall computation exponential in $n$.}
Nevertheless, OASIS-GT provides a principled foundation for operator-aware importance sampling, and enhancing the scalability of the underlying optimization remains an important challenge.

While \hl{OASIS-GHZ/W} demonstrate both strong performance and scalability, it still inherits a limitation in that it is currently applicable only to Pauli measurement settings. Extending the framework to accommodate more general measurement families remains an important direction for future work.

Additional promising directions include enhancing OASIS-GT through improved surrogate formulations and by considering other IOC-POVMs, and developing \hl{target-specific algorithms} beyond the GHZ and W states.

\begin{acknowledgments}
This work is in part supported by the National Research Foundation of Korea (NRF, RS-2024-00451435 (20\%), RS-2024-00413957 (20\%)), Institute of Information \& communications Technology Planning \& Evaluation (IITP, RS-2021-II212068 (10\%), RS-2025-02305453 (15\%), RS-2025-02273157 (15\%), RS-2025-25442149 (10\%), RS-2021-II211343(10\%)), a grant funded by the Ministry of Science and ICT (MSIT), Institute of New Media and Communications (INMAC), and the BK21 FOUR program of the Education, Artificial Intelligence Graduate School Program (Seoul National University), and Research Program for Future ICT Pioneers, Seoul National University in 2025.
\end{acknowledgments}

\section*{Data availability}

The data that support the findings of this article are openly available~\cite{cha2025oasiscode}.

\bibliography{apssampv1}

\clearpage

\appendix

\section{The sorted insertion algorithm}
\label{appendix:sorted_insertion}

The following algorithm describes the SI method for grouping Pauli operators to reduce estimator variance.

\begin{figure}[H]
\centering
\begin{minipage}{\linewidth}
\caption{\justifying Sorted insertion \cite{crawford2021efficient}.}
\label{algorithm:sorted_insertion_algorithm}
\rule{\linewidth}{1pt}
\begin{algorithmic}[0]
\State \textbf{Input:} pauli\_list and chi\_list, where
\State \(\text{chi\_list[i]} = \chi_O(\text{pauli\_list[i]})\)
\State \textbf{Output:} pauli\_groups and chi\_groups
\State Sort pauli\_list and chi\_list in decreasing order of the
\State absolute values of the coefficients in chi\_list.
\State pauli\_groups, chi\_groups \(\gets\) [], []
\For{i in range(\(d^2\))}
    \State inserted\(\,\gets\,\)False
    \For{j in range(len(pauli\_groups))}
        \If{pauli\_list[i] commutes with pauli\_groups[j]}
            \State pauli\_groups[j].append(pauli\_list[i])
            \State chi\_groups[j].append(chi\_list[i])
            \State inserted\(\,\gets\,\)True
            \State \textbf{break}
        \EndIf
    \EndFor
    \If{not inserted}
        \State pauli\_groups.append([pauli\_list[i]])
        \State chi\_groups.append([chi\_list[i]])
    \EndIf
\EndFor
\State \textbf{return} pauli\_groups, chi\_groups
\end{algorithmic}
\rule{\linewidth}{1pt}
\end{minipage}
\end{figure}

\section{Derivation of Algorithm~\ref{algorithm:oasis_lp}}
\label{appendix:algorithm_1_derivation}

Suppose a given measurement scheme yields an IOC-POVM
\[
\boldsymbol\Pi = \{\Pi_{U,\boldb}\}_{U,\boldb},
\]
where \(U\) denotes an \(n\)-qubit unitary and \(\boldb\in\{0,1\}^n\) represents a measured outcome, with the completeness relation
\[
\sum_{U,\boldb} \Pi_{U,\boldb}=I.
\]
Let \(p(U)\) denote the \emph{default} distribution of \(U\) that defines \(\boldsymbol\Pi\). In other words, we apply \(\rho \mapsto U\rho U^\dagger\) with probability \(p(U)\). Then the POVM elements are
\[
\Pi_{U,\boldb} = p(U)U^\dagger|\boldb\rangle\langle\boldb|U.
\]
Also, let \(f(\boldb;\rho,U)\) be the probability of measuring \(\boldb\) when \(U\rho U^\dagger\) is measured in the computational basis. Then \(f(\boldb;\rho,U)\) can be written as
\[
f(\boldb;\rho,U)=\frac{\tr(\rho \Pi_{U,\boldb})}{p(U)}.
\]
Let us denote the target state's density matrix as \(O\). Then our goal can be formulated as estimating \(\tr(\rho O)\) given many copies of an unknown state \(\rho\). Since \(\boldsymbol\Pi\) is an IOC-POVM, \(O\) can be expressed as a linear combination of the elements of \(\boldsymbol\Pi\) as
\[
O = \sum_{U,\boldb} \omega_{U,\boldb} \Pi_{U,\boldb},
\]
where the weights \(\omega_{U,\boldb}\in\R\) may not be uniquely determined \cite{d2004informationally, zhu2014quantum, innocenti2023shadow, fischer2024dual, malmi2024enhanced}.

Suppose now that \(U\) is sampled from another distribution \(q(U)\), satisfying \(\sum_U q(U)=1\), instead of \(p(U)\). After sampling \(U \sim q(U)\) and obtaining outcome \(\boldb\) from measuring \(U\rho U^\dagger\) in the computational basis, we define the estimator $S(U,\boldb)$ as
\[
S(U,\boldb) = \frac{\omega_{U,\boldb} p(U)}{q(U)}.
\]
With this definition, the estimator is unbiased, as its expectation satisfies
\begin{align*}
\E_{U,\boldb}[S(U,\boldb)] & = \sum_U q(U) \sum_\boldb f_\boldp(\boldb;\rho,U) \frac{\omega_{U,\boldb} p(U)}{q(U)}\\
& = \sum_{U,\boldb} \tr(\rho \Pi_{U,\boldb}) \omega_{U,\boldb}\\
& = \tr(\rho O).
\end{align*}
Then we can minimize \(\text{Var}(S)\) by minimizing \(\E_{U,\boldb}[S(U,\boldb)^2]\). However, this quantity is unknown, as it depends on the measured state \(\rho\). Therefore, we propose minimizing the \emph{surrogate}
\begin{align}
\label{equation:surrogate}
\E_U\!\left[\max_\boldb S(U,\boldb)^2\right] & = \sum_U q(U) \max_\boldb \frac{\omega_{U,\boldb}^2 p(U)^2}{q(U)^2}\\
& = \sum_U \frac{p(U)^2}{q(U)} \max_\boldb \omega_{U,\boldb}^2,\nonumber
\end{align}
which upper bounds \(\text{Var}(S)\), optimized over \((\omega,q)\). Observe that, for fixed \(\omega\), the distribution \(q\) that minimizes Eq.~\eqref{equation:surrogate} is given by
\[
q(U) \propto p(U) \max_\boldb |\omega_{U,\boldb}|,
\]
in which case Eq.~\eqref{equation:surrogate} evaluates to
\[
\left(\sum_U p(U) \max_\boldb |\omega_{U,\boldb}|\right)^2.
\]
The optimization problem can therefore be summarized as follows:
\begin{align}
\mathop{\text{minimize}}_{\omega} & \quad \sum_U p(U) \max_\boldb |\omega_{U,\boldb}|\\
\text{subject to} & \quad \sum_{U,\boldb} \omega_{U,\boldb} \Pi_{U,\boldb} = O.\nonumber
\end{align}
Equivalently, we can solve the following LP:
\begin{align}
\mathop{\text{minimize}}_{\omega,t} & \quad \sum_U p(U) t_U\\
\text{subject to} & \quad -t_U \le \omega_{U,\boldb} \le t_U \quad \forall U,\boldb,\nonumber\\
& \quad \sum_{U,\boldb} \omega_{U,\boldb} \Pi_{U,\boldb} = O.\nonumber
\end{align}
After optimizing \(\omega\), we sample \(U\) from the distribution
\[
q(U) = \frac{p(U) \max_\boldb |\omega_{U,\boldb}|}{\sum_{U^\prime} p(U^\prime) \max_\boldb |\omega_{U^\prime,\boldb}|}.
\]

\begin{remark}
\label{remark:why_lp_worse}
For a Haar-random state, with probability one the Pauli expansion is essentially dense, i.e., almost all Pauli coefficients are nonzero. As a result, both G-DFE and OASIS-GT effectively must sample from (nearly) the full set of \(3^n\) local Pauli measurement settings. In this regime, OASIS-GT can outperform by globally optimizing its estimator to reduce a variance upper bound rather than relying on G-DFE's greedy grouping. In contrast, for highly structured targets such as GHZ and W, most Pauli coefficients are exactly zero and the nonzero support is concentrated in a small subset. G-DFE directly exploits this sparsity by focusing only on the nonzero-probability terms and grouping them efficiently. OASIS-GT, however, optimizes a worst-case variance bound over an overcomplete POVM expansion, which need not assign zero weight to many irrelevant Pauli strings. Consequently, it may continue to spread effort over measurement settings that are redundant for these structured targets, and can therefore fail to outperform (or even underperform) G-DFE in this setting.
\end{remark}

\begin{remark}
Both G-DFE and OASIS-GT incur $2^{\Theta(n)}$ offline (constructing the fidelity estimator) cost for general target states.
However, we highlight that the practical utility of our approach lies in the online (the actual fidelity estimation) phase of repeated experimental implementations.
In G-DFE, each shot samples a group and produces a scalar contribution that involves computing a weighted sum over the eigenvalues associated with the sampled group \cite{barbera2025sampling}.
The paper does not explicitly isolate the per-shot classical time of this computation, but it is clear that the online classical processing entails a nontrivial group-dependent weighted sum (whose evaluation scales with the group size).
In contrast, OASIS-GT returns the estimator in its most direct and deployment-friendly form.
Once the optimization produces $q(U)$ and the estimator $S(U,\mathbf{b})$, the online procedure can be implemented with an alias table~\cite{walker1974new, walker1977efficient, vose1991linear} for sampling $U\sim q(U)$ and a lookup table for $S(U,\mathbf{b})$.
Consequently, the online classical compute per shot is essentially $O(1)$.
This difference is particularly beneficial in the common regime where the same optimized fidelity estimator is reused arbitrarily many times.
In such cases, the offline cost is amortized, and one repeatedly gains not only the primary benefit of reduced variance in the online fidelity estimate, but also a potentially lower per-measurement classical processing overhead.
\end{remark}

\section{Derivation of Algorithm~\ref{algorithm:oasis_st_ghz}}
\label{appendix:algorithm_2_derivation}

We treat the identity string as a singleton group (branch 0 in Algorithm~\ref{algorithm:oasis_st_ghz}), since its expectation value is always 1 and thus requires no copies of \(\rho\) for estimation.

For the GHZ state, the probabilities of Pauli strings defined in Eq.~\eqref{equation:dfe_probabilities} are given by
\begin{equation*}
\tilde{p}(\boldp) = \begin{cases}
    \frac{1}{d} & \text{if } \boldp\in\{I,Z\}^n \text{ and } |\boldp_Z| \equiv 0 \, (\mathrm{mod} \, 2)\\
    \frac{1}{d} & \text{if } \boldp\in\{X,Y\}^n \text{ and } |\boldp_Y| \equiv 0 \, (\mathrm{mod} \, 2)\\
    0 & \text{otherwise}
\end{cases}.
\end{equation*}
We highlight the following properties:
\begin{enumerate}
    \item Each \(\boldp\in\{I,Z\}^n\) qubit-wise commutes with the pivot \(\boldp=Z\cdots Z\), which we denote by \(\boldp^{(Z)}\).
    \item Each \(\boldp\in\{X,Y\}^n\) with \(|\boldp_Y| \equiv 0 \, (\mathrm{mod} \, 2)\) does not commute with any other non-identity string \(\boldp\) satisfying \(\tilde{p}(\boldp)>0\). Therefore, such \(\boldp\) is itself a pivot and forms a group on its own.
\end{enumerate}
In other words, each pivot serves as a representative Pauli string for its group.
The corresponding group probabilities are:
\begin{enumerate}
    \item \(\boldp^{(Z)}\): The group contains \(\left(\frac{d}{2}-1\right)\) Pauli strings (except for \(I\cdots I\)). Therefore, the group probability is \(\left(\frac{d}{2}-1\right)\frac{1}{d}=\frac{d-2}{2d}\).

    \item \(\boldp\in\{X,Y\}^n\) with \(|\boldp_Y| \equiv 0 \, (\mathrm{mod} \, 2)\): The group probability is \(1/d\).
\end{enumerate}

Note that
\begin{align*}
& \sum_{\substack{\boldv \in \{I,Z\}^n,\\ |\boldv_Z|>0,\\ |\boldv_Z|\equiv0\,(\mathrm{mod}\,2)}} \tr\!\left( \rho \bigotimes_{i=1}^n \boldv_i \right)\\
= & \sum_{\substack{\boldv \in \{I,Z\}^n,\\ |\boldv_Z|\equiv0\,(\mathrm{mod}\,2)}} \sum_{\boldb\in\{0,1\}^n} \langle\boldb|\rho|\boldb\rangle (-1)^{\boldv_Z \cdot \boldb} - 1\\
= & \E_\boldb\!\left[ \sum_{\substack{\boldv \in \{I,Z\}^n,\\ |\boldv_Z|\equiv0\,(\mathrm{mod}\,2)}} (-1)^{\boldv_Z \cdot \boldb} \right] - 1,
\end{align*}
where the expectation is taken over the measurement outcome \(\boldb\) obtained from measuring \(\rho\) in the computational basis. Moreover,
\begin{align*}
& \sum_{\substack{\boldv \in \{I,Z\}^n,\\ |\boldv_Z|\equiv0\,(\mathrm{mod}\,2)}} (-1)^{\boldv_Z \cdot \boldb}\\
= & \frac{1}{2} \sum_{\boldv \in \{I,Z\}^n} (-1)^{\boldv_Z \cdot \boldb} + \frac{1}{2} \sum_{\boldv \in \{I,Z\}^n} (-1)^{|\boldv_Z|} (-1)^{\boldv_Z \cdot \boldb}\\
= & \frac{1}{2} \prod_{i=1}^n \left(1+(-1)^{\boldb_i}\right) + \frac{1}{2} \prod_{i=1}^n \left(1-(-1)^{\boldb_i}\right)\\
= & \frac{d}{2} (\delta_{\boldb,\mathbf{0}}+\delta_{\boldb,\mathbf{1}}).
\end{align*}
Therefore, if \(\boldp = \boldp^{(Z)}\) (branch 1) and the outcome \(\boldb\in\{0,1\}^n\) is observed, then one can verify that
\begin{align}
\label{equation:ghz_brahch1_S}
S(\boldp,\boldb) & = \frac{2d}{d-2} \frac{1}{\sqrt{d}} \frac{(d/2)(\delta_{\boldb,\mathbf{0}}+\delta_{\boldb,\mathbf{1}}) - 1}{\sqrt{d}}\nonumber\\
& = \begin{cases}
    1 & \text{if } |\boldb|=0 \text{ or } |\boldb|=n\\
    -\frac{2}{d-2} & \text{otherwise}
\end{cases}
\end{align}
is an unbiased estimator of \(R\) in Eq.~\eqref{equation:barbera_random_variable_definition}.

If \(\boldp\in\{X,Y\}^n\), \(|\boldp_Y| \equiv 0 \, (\mathrm{mod} \, 2)\) (branch 2), and the outcome \(\boldb\in\{0,1\}^n\) is observed, then one can verify that
\begin{equation}
\label{equation:ghz_brahch2_S}
S(\boldp,\boldb) = d \frac{(-1)^{|\boldp_Y|/2}}{\sqrt{d}} \frac{(-1)^{|\boldb|}}{\sqrt{d}} = (-1)^{|\boldp_Y|/2 + |\boldb|}
\end{equation}
is an unbiased estimator of \(R\) in Eq.~\eqref{equation:barbera_random_variable_definition}, because
\begin{align*}
\tr\!\left(O \bigotimes_{i=1}^n \boldp_i\right) & = \frac{1}{2} \left( \langle\mathbf{0}| \bigotimes_{i=1}^n \boldp_i |\mathbf{1}\rangle + \langle\mathbf{1}| \bigotimes_{i=1}^n \boldp_i |\mathbf{0}\rangle \right)\\
& = \frac{1}{2} \left( \langle\mathbf{0}| (-i)^{|\boldp_Y|} |\mathbf{0}\rangle + \langle\mathbf{1}| i^{|\boldp_Y|} |\mathbf{1}\rangle \right)\\
& = (-1)^{|\boldp_Y|/2}.
\end{align*}

Therefore, the fidelity can be written as
\begin{equation}
\label{equation:s1s2_expression_ghz}
\frac{1}{d} + \frac{d-2}{2d}\E[S_1] + \frac{1}{2}\E[S_2],
\end{equation}
where
\begin{align}
\label{equation:s1s2_definition}
& S_1 = (S(\boldp,\boldb) \mid \text{branch 1})\nonumber\\
\text{and} \quad & S_2 = (S(\boldp,\boldb) \mid \text{branch 2}).
\end{align}

Lastly, we calculate the number of copies required to estimate \(R\) using Eq.~\eqref{equation:ghz_brahch1_S} or Eq.~\eqref{equation:ghz_brahch2_S}. It is given by
\begin{equation}
\label{equation:group_num_shots_formula}
m_\boldp = \left\lceil \frac{2\left\|(\boldsymbol\chi_O)_{\mathcal{I}(\boldp)}\right\|_1^2}{\left\|(\boldsymbol\chi_O)_{\mathcal{I}(\boldp)}\right\|^4 d l \epsilon^2} \ln\frac{2}{\delta} \right\rceil
\end{equation}
\cite{barbera2025sampling}. In branch 1,
\begin{equation*}
m_\boldp = \left\lceil \frac{2\left(\frac{1}{\sqrt{d}}\left(\frac{d}{2}-1\right)\right)^2}{\left(\frac{1}{d}\left(\frac{d}{2}-1\right)\right)^2 dl\epsilon^2} \ln\frac{2}{\delta} \right\rceil = \left\lceil \frac{2}{l\epsilon^2} \ln\frac{2}{\delta} \right\rceil.
\end{equation*}
In branch 2, \((\boldsymbol\chi_O)_{\mathcal{I}(\boldp)} \in \R^1\) and \(\left|(\boldsymbol\chi_O)_{\mathcal{I}(\boldp)}\right| = 1 / \sqrt{d}\), so
\[
m_\boldp = \left\lceil \frac{2}{dl\epsilon^2} \frac{1}{d} d^2 \ln\frac{2}{\delta} \right\rceil = \left\lceil \frac{2}{l\epsilon^2} \ln\frac{2}{\delta} \right\rceil.
\]

\begin{remark}
\label{remark:ghz_not_scalable}
When the identity string is isolated, it can be shown that G-DFE for the GHZ state reproduces exactly the grouping described in Algorithm~\ref{algorithm:oasis_st_ghz}. Since all Pauli strings with nonzero probabilities occur with equal probability, any ordering is admissible when sorting them in descending order in the SI algorithm. However, upon inserting \(\boldp \in \{I,Z\}^n\), it can only be grouped with strings in \(\{I,Z\}^n\). Similarly, when inserting a string \(\boldp \in \{X,Y\}^n\), it cannot be merged with any existing group. Consequently, all strings in \(\{I,Z\}^n\) are eventually grouped together under the \emph{pivot} (i.e., the representative measurement setting for this group) \(\boldp^{(Z)}\), while each string in \(\{X,Y\}^n\) forms a singleton group. Moreover, this grouping is optimal in the sense that the number of groups cannot be further reduced.
While the grouping coincides with that obtained by a vanilla implementation of G-DFE, the computational implications are vastly different. The original G-DFE algorithm would require exponential time and memory to enumerate and store all groups, whereas our formulation provides a compact description that enables the same grouping to be realized efficiently.
\end{remark}

\section{Variance of the estimator in Algorithm~\ref{algorithm:oasis_st_ghz}}
\label{appendix:algorithm_2_variance}

Suppose \(d\ge 4\). Continuing from the definition in Eq.~\eqref{equation:s1s2_definition}, let
\begin{align*}
\E[S_1] & =\frac{pd-2}{d-2}, & \mathrm{Var}(S_1) & =\left(\frac{d}{d-2}\right)^2 p(1-p),\\
\E[S_2] & =2q-1, & \mathrm{Var}(S_2) & =4q(1-q),
\end{align*}
and \(f=\tr(\rho O)\). From Eq.~\eqref{equation:s1s2_expression_ghz}, we have
\begin{equation}
\label{equation:p_and_q_relation}
\frac{pd-2}{2d} + \frac{2q-1}{2} = f-\frac{1}{d} \quad \Longrightarrow \quad p=2f+1-2q.
\end{equation}
Using the law of total variance, in Algorithm~\ref{algorithm:oasis_st_ghz},
\begin{align*}
\mathrm{Var}(S) & = \frac{(d-2)\mathrm{Var}(S_1) + d\mathrm{Var}(S_2)}{2d}\\
& + \frac{(\E[S_1]-\E[S_2])^2}{4}\\
& + \frac{\E[S_1]-1}{d}\left(\E[S_2]-1-\frac{\E[S_1]-1}{d}\right).
\end{align*}
A short algebraic simplification gives
\[
\mathrm{Var}(S) = 1 - f^2 - (q-f)\frac{d-4}{d-2}.
\]
Feasibility requires \(0\le p\le 1\) and \(0\le q\le 1\), which together with the constraint Eq.~\eqref{equation:p_and_q_relation} imply \(q\ge f\). Hence
\[
\mathrm{Var}(S) \le 1-f^2.
\]
This bound is tight, attained at \(p=1\) and \(q=f\). In particular, \(\mathrm{Var}(S) \rightarrow 0\) as \(f \rightarrow 1\).

\section{Derivation of Algorithm~\ref{algorithm:oasis_st_w}}
\label{appendix:algorithm_3_derivation}

As in Appendix~\ref{appendix:algorithm_2_derivation}, we assume that the identity string is treated separately (branch 0 in Algorithm~\ref{algorithm:oasis_st_w}).

For the W state, the probabilities of Pauli strings defined in Eq.~\eqref{equation:dfe_probabilities} are given by
\begin{equation*}
\tilde{p}(\boldp) = \begin{cases}
    \frac{(n-2|\boldp_Z|)^2}{n^2 d} & \text{if } \boldp\in\{I,Z\}^n\\
    \frac{4}{n^2 d} & \text{if } |\boldp_X|=2 \text{ and } \boldp\in\{I,X,Z\}^n\\
    \frac{4}{n^2 d} & \text{if } |\boldp_Y|=2 \text{ and } \boldp\in\{I,Y,Z\}^n\\
    0 & \text{otherwise}
\end{cases}.
\end{equation*}
We highlight the following properties:
\begin{enumerate}
    \item Each \(\boldp\in\{I,Z\}^n\) qubit-wise commutes with the pivot \(\boldp^{(Z)}\).
    \item Each \(\boldp\in\{I,X,Z\}^n\) with \(\boldp_i=\boldp_j=X\) and \(\boldp_{k\notin\{i,j\}}\in\{I,Z\}\) qubit-wise commutes with the pivot \(\boldp\in\{X,Z\}^n\) with \(\boldp_i=\boldp_j=X\) and \(\boldp_{k\notin\{i,j\}}=Z\), which we denote by \(\boldp^{(X,i,j)}\).
    \item Each \(\boldp\in\{I,Y,Z\}^n\) with \(\boldp_i=\boldp_j=Y\) and \(\boldp_{k\notin\{i,j\}}\in\{I,Z\}\) qubit-wise commutes with the pivot \(\boldp\in\{Y,Z\}^n\) with \(\boldp_i=\boldp_j=Y\) and \(\boldp_{k\notin\{i,j\}}=Z\), which we denote by \(\boldp^{(Y,i,j)}\).
\end{enumerate}
The corresponding group probabilities are:
\begin{enumerate}
    \item \(\boldp^{(X,i,j)}\): The group contains \(d/4\) Pauli strings. Therefore, the group probability is \(\tilde{p}\!\left(\boldp^{(X,i,j)}\right) = \frac{d}{4} \frac{4}{n^2 d} = \frac{1}{n^2}\).

    \item \(\boldp^{(Y,i,j)}\): Similarly, the group probability is \(\tilde{p}\!\left(\boldp^{(Y,i,j)}\right) = \frac{1}{n^2}\).

    \item \(\boldp^{(Z)}\): The group probability (except for \(I\cdots I\)) can be obtained by subtracting the contributions of \(\boldp^{(X,i,j)}\), \(\boldp^{(Y,i,j)}\), and \(I\cdots I\) from 1: \(\tilde{p}\!\left(\boldp^{(Z)}\right) = 1-2\binom{n}{2}\frac{1}{n^2}-\frac{1}{d}=\frac{d-n}{nd}\).
\end{enumerate}

If \(\boldp = \boldp^{(Z)}\) (branch 1) and the outcome \(\boldb\in\{0,1\}^n\) is observed, then one can verify that
\[
S(\boldp,\boldb) = \frac{1}{d-n} \left( \sum_{\boldv\in\{I,Z\}^n} (n - 2|\boldv_Z|) (-1)^{\boldb \cdot \boldv_Z} - n \right)
\]
is an unbiased estimator of \(R\) in Eq.~\eqref{equation:barbera_random_variable_definition}. But since
\begin{align*}
& \sum_{\boldv\in\{I,Z\}^n} (n - 2|\boldv_Z|) (-1)^{\boldb \cdot \boldv_Z}\\
= & \sum_{\boldv\in\{I,Z\}^n} \sum_{i=1}^n (1 - 2 (\boldv_Z)_i) (-1)^{\boldb \cdot \boldv_Z}\\
= & \sum_{\boldv\in\{I,Z\}^n} \sum_{i=1}^n (-1)^{(\boldv_Z)_i} (-1)^{\boldb \cdot \boldv_Z}\\
= & \sum_{i=1}^n \sum_{\boldv\in\{I,Z\}^n} (-1)^{(\boldb + \mathbf{e}_i) \cdot \boldv_Z}\\
= & \sum_{i=1}^n d \mathbf{1}_{\{\mathbf{e}_i\}}(\boldb)\\
= & \begin{cases}
    d & \text{if } |\boldb|=1\\
    0 & \text{otherwise}
\end{cases},
\end{align*}
where \(\mathbf{e}_i\) denotes the \(i\)-th standard basis vector, we have
\begin{equation}
\label{equation:w_brahch1_S}
S(\boldp,\boldb) = \begin{cases}
    1 & \text{if } |\boldb|=1\\
    -\frac{n}{d-n} & \text{otherwise}
\end{cases}.
\end{equation}

If \(\boldp = \boldp^{(X/Y,i,j)}\) (branch 2) and the outcome \(\boldb\in\{0,1\}^n\) is observed, then one can verify that
\begin{align}
\label{equation:w_brahch2_S}
S(\boldp,\boldb) & = \frac{n^2}{\sqrt{d}} (-1)^{b_i + b_j} \sum_{\substack{\boldv\in\{I,X/Y,Z\}^n,\\ |\boldv_{X/Y}|=2,\\ \boldv_i=\boldv_i=X/Y}} \frac{2}{n\sqrt{d}} (-1)^{\boldb \cdot \boldv_Z}\nonumber\\
 & = \begin{cases}
     \frac{n}{2} (-1)^{b_i + b_j} & \text{if } \left|\boldb_{[n]\setminus\{i,j\}}\right|=0\\
     0 & \text{otherwise}
 \end{cases}
\end{align}
is an unbiased estimator of \(R\) in Eq.~\eqref{equation:barbera_random_variable_definition}.

Therefore, the fidelity can be written as
\begin{equation*}
\frac{1}{d} + \frac{d-n}{nd}\E[S_1] + \frac{n-1}{n}\E[S_2],
\end{equation*}
where \(S_1\) and \(S_2\) are defined in Eq.~\eqref{equation:s1s2_definition}.

Lastly, we calculate the number of copies required to estimate \(R\) using Eqs.~\eqref{equation:group_num_shots_formula},~\eqref{equation:w_brahch1_S}, and~\eqref{equation:w_brahch2_S}. In branch 1,
\begin{align*}
m_\boldp & = \left\lceil \frac{2\left(\sum_{i=0}^n\binom{n}{i}\frac{|n-2i|}{n\sqrt{d}} - \frac{1}{\sqrt{d}}\right)^2}{dl\epsilon^2} \frac{n^2 d^2}{(d-n)^2} \ln\frac{2}{\delta} \right\rceil\\
& = \left\lceil \frac{2\left(\sum_{i=0}^n\binom{n}{i}|n-2i| - n\right)^2}{l\epsilon^2 (d-n)^2} \ln\frac{2}{\delta} \right\rceil\\
& = \left\lceil \frac{2\left(2n\binom{n-1}{\lfloor n/2\rfloor} - n\right)^2}{l\epsilon^2 (d-n)^2} \ln\frac{2}{\delta} \right\rceil\\
& = \left\lceil \frac{2 n^2}{l\epsilon^2} \left(\frac{2\binom{n-1}{\lfloor n/2\rfloor} - 1}{d-n}\right)^2 \ln\frac{2}{\delta} \right\rceil.
\end{align*}
In branch 2, \(|\mathcal{I}(\boldp)| = d/4\) and all entries of \((\boldsymbol\chi_O)_{\mathcal{I}(\boldp)}\) are \(2 / n\sqrt{d}\). Therefore,
\[
m_\boldp = \left\lceil \frac{2}{dl\epsilon^2} \frac{d}{4n^2} n^4 \ln\frac{2}{\delta} \right\rceil = \left\lceil \frac{n^2}{2l\epsilon^2} \ln\frac{2}{\delta} \right\rceil.
\]

If the identity string is isolated, then the number of groups cannot be reduced beyond the grouping we described because all other pivots are mutually non-commuting. Analogous to Algorithm~\ref{algorithm:oasis_st_ghz}, our estimator for the W state removes the exponential resource overhead of G-DFE.

\section{Additional tables}
\label{section:additional_tables}

The following tables provide additional numerical data \hl{for our experiments}.

\begin{table}[h]
\caption{\label{table:num_shots}\justifying Average number of measurement shots used by G-DFE.}
\centering
\begin{tabular}{l|cccc}
\toprule
\multicolumn{1}{c|}{$n$} & 3 & 4 & 5 & 6\\ 
\midrule
Haar & 4426.0 & 8126.6 & 14083.3 & 27399.4\\ \hline
GHZ & 875.6 & 937.5 & 968.6 & 984.3\\ \hline
W & 1749.8 & 2625.1 & 3707.1 & 5453.5\\
\bottomrule
\end{tabular}
\end{table}

\begin{table}[h]
\caption{\label{table:num_groups}\justifying Number of groups produced by G-DFE, \hl{OASIS-GHZ, and OASIS-W}.}
\centering
\begin{tabular}{l|l|cccc}
\toprule
& \multicolumn{1}{c|}{$n$} & 3 & 4 & 5 & 6\\
\midrule
\multirow{2}{*}{GHZ} & G-DFE & 6 & 10 & 18 & 34\\
                     & OASIS-GHZ & 6 & 10 & 18 & 34\\
\hline
\multirow{2}{*}{W}   & G-DFE & 8 & 16 & 26 & 35\\
                     & OASIS-W & 8 & 14 & 22 & 32\\
\bottomrule
\end{tabular}
\end{table}

\end{document}